# E- Exams System for Nigerian Universities with Emphasis on Security and Result Integrity

**Olawale Adebayo M.Sc.**

Department of Cyber and Security Science, Federal University of Technology Minna, Niger State, Nigeria.

E-Mail: waleadebayo@futminna.edu.ng1

**Shafi'i Muhammad Abdulhamid M.Sc.**

Department of Cyber and Security Science, Federal University of Technology Minna, Niger State, Nigeria.

E-Mail: shafzon@yahoo.com, shafii.abdulhamid@futminna.edu.ng

*Abstract*- **The recent employment and eventual widespread acceptance of electronic test in examining students and various classes in Nigeria has created a significant impact in the trends of educational history in the country. In this paper, we examined the impacts, associated challenges and security lapses of the existing electronic-examination system with the aim of ameliorating and developing a new acceptable e-Exam system that takes care of the existing system's challenges and security lapses. Six Universities that are already conducting e-Examination were selected across the country for this research work. Twenty (20) students that participated in the e-exams and five (5) members of staff were selected for interview and questionnaire. Based on the analysis of the interviews and study of the existing electronic examination system, some anomalies were discovered and a new e-exams system was developed to eradicate these anomalies. The new system uses data encryption in order to protect the questions sent to the e-Examination center through the internet or intranet and a biometric fingerprint authentication to screen the stakeholders.**

*Keywords*- **Biometric Fingerprint, Cryptography, eLearning, Electronic Examination (e-exams), Result Integrity**

## INTRODUCTION

Electronic examination has been highly interested and suitable in both educational and pedagogical aspects. Examination is one of the best methods of evaluating the knowledge and ability of an individual. To this end, various methods has been employed in examining the ability of an individual, starting from manual means of using paper and pencil to electronic, from oral to written, practical to theoretical and many others.

The present information technology means of examining students is the use of electronic systems in place of manual or paper method which was characterized by massive examination leakages, impersonations, demand for gratification by teachers, bribe-taking by supervisors and invigilators of examinations.

The employers are conducting aptitude test for their job seekers through electronic means; the universities and other tertiary institutions are registering and conducting electronic examination for their students through the internet and other electronic and networking gadgets, various examination bodies in the country like the West Africa Examination Council (WAEC), National Examination Council (NECO), National Board for Technical Education (NABTEB), National Teacher Institute (NTI) e.t.c. register their students through electronic means, recently electronic examination has been widely adopted by nearly all the Nigeria University for post Unified Tertiary and Matriculation Examination (Post-UTME) otherwise called pre-admission screening. With these aforementioned and many more educational bodies engaging in electronic examination and registration for testing the ability of their candidates, which determine the future of this great country and our dear youth, there is need for serious examination of the system which has great impacts on the populace.

## LITERATURE REVIEW

There is a growing body of research focused on developing better ways to manage e-exams systems and e-learning systems. Some of these researches focused on various sections of the system and these includes:

Schramm (2008) looked at a e-learning web based system that could simply offer and grade mathematical questions with infinite patience. Therefore it needs the capability for in and output of mathematical formulas, the dynamic generation of plots and the generation of random expressions and numbers. Al-Bayati and Hussein (2008) presents an applied Generic Software of multiple kinds of e-exam package; this package of e-exam is oriented to Hearing Impaired (HI) persons. Therefore the exam material of this package is translated into language of HI persons like sign language and finger spelling. The idea of the Generic software is to present an empty templates to the teacher who would like to develop his required e-exam for the needful topic (mathematics, language, science, etc) and desired set of exam kinds (multiple choices, matching between words, fill in blanks, etc).

Web-based Examination System is an effective solution for mass education evaluation (Zhenming et al, 2003). They developed a novel online examination system based on a Browser/Server framework which carries out the examination and auto-grading for objective questions and operating questions, such as programming, operating Microsoft Windows, editing Microsoft Word , Excel and PowerPoint, etc. It has been successfully applied to the distance evaluation of basic operating skills of computer science, such as the course of computer skills in Universities and the nationwide examination for the high school graduates in Zhejiang Province, China. Another paper (He, 2006) presents a web-based educational assessment system by applying Bloom's taxonomy to evaluate student learning outcomes and teacher instructional practices in real time. The system performance is rather encouraging with experimentation in science and mathematics courses of two local high schools.

Another paper proposed web based online examination system (Rashad et al, 2010). The system carries out the examination and auto-grading for students exams. The system facilitates conducting exams, collection of answers, auto marking the submissions and production of reports for the test. It supports many kinds of questions. It was used via Internet and is therefore suitable for both local and remote examination. The system could help lecturers, instructors, teachers and others who are willing to create new exams or edit existing ones as well as students participating in the exams. The system was built using. various open source technologies AJAX, PHP, HTML and MYSQL database are used in this system. An auto-grading module was generalized to enable different exam and question types. The system was tested in the Mansoura university quality assurance center. The test proved the validity of using this kind of web based systems for evaluates students in the institutions with high rate of students.

An online website for tutoring and e-examination of economic course aimed to present a novel software tool can be used for online examination and tutorial application of the syllabus of economic course (EL Emary and Al Sondos, 2006). Also, among the main interests of the paper is to produce a software through it we make sure that students have studied all the concepts of economics. So, the proposed software is structured from two major modules: The first one was an online website to review and make self-test for all the material of economic course. The second part is an online examination using a large database bank of questions through it the level of students can be evaluated immediately and some statistical evaluations can be obtained. The developed software offers the following features: 1) Instructors could add any further questions to maximize the size of the bank of questions. 2) Different examinations for each student with randomly selected questions from the bank of questions can be done. 3) Different reports for the instructors, students, classes…etc can be obtained. 4) Several students can take their exams simultaneously without any problem inside and outside their campus. The proposed software has been designed to work base on the client server architecture.

Electronic exam is a difficult part of e-learning security (Huszti and Petho, 2008). The paper describes a cryptographic scheme that possesses security requirements, such that authenticity, anonymity, secrecy, robustness, correctness without the existence of a Trusted Third Party. The proposed protocol also provides students a receipt, a proof of a successful submission, and it is based on existence of anonymous return channels. Another research work proposed a theoretical approach that incorporates available fingerprint biometrics authentication technologies in conjunction with e-learning environments to curb unethical conduct during e-learning exam taking (Levy and Ramim, 2007). The proposed approach suggests practical solution that can incorporate a random fingerprint biometrics user authentication during exam taking in e-learning courses. Doing so is hypothesized to curb exam cheating in e-learning environments.

Ayo et al (2007) proposed a model for e-Examination in Nigeria where all applicants are subjected to online entrance examination as a way of curbing the irregularities as proposed by the Joint Admissions Matriculation Board (JAMB), the body saddled with the responsibility of conducting entrance examinations into all the Nigerian universities. This model was designed and tested in Covenant University, one of the private universities in Nigeria. Their findings revealed that the system has the potentials to eliminate some of the problems that are associated with the traditional methods of examination such as impersonation and other forms of examination malpractices. Based on the development of e-learning in the only Open University in Nigeria (Ipaye, 2009), discusses the process of establishing e-learning environment. Another paper seeks to solve a part of that problem by designing and developing a web application where tests in multiple choice formats will be taken online and graded immediately (Akinsanmi et al, 2010). The web application relies solely on Microsoft developed technologies. It runs on the Microsoft.net framework, uses the ASP.NET web server, C# as the intermediate language, ADO.NET to interact with the relational database and Microsoft SQL server as the relational database.

**ANALYSIS OF THE EXISTING SYSTEMS USED IN NIGERIA**

In Nigeria, very few Universities have started using the e-exams system for their test/exams and these includes Federal University of Technology Minna, University of Ilorin, Covenant University Ota, Nigerian Open University of Nigeria (NOUN), to mention but a few. In all the six Universities visited in the course of this research, they are all operating almost in the same way. Only NOUN uses internet for the e-exams, while others uses intranet setup within the University environments. The intranet was setup in e-exams centers containing 50 to 200 computer systems and a server. Another observation made was that most of these centers are being managed by private company (Electronic Test Company Limited) which is not so good for the integrity of the results.

**Architecture of the Existing System**

Ayo et al (2007) and Akinsanmi (2010) presented a 3-tier architecture comprising the presentation tier, the logic tier and the database tier. The presentation tier offers an interface to the user, the logic tier serves as the middleware that is responsible for processing the user's requests, while the database tier serves as the repository of a pool of thousands of questions. It also consists of other modules for authentication (using User name/Registration Number and Password) and computing results. This is the architecture used by all the e-exams centers visited within Nigeria and it is also the same architecture that was used even in other countries with just little modifications. This type of architecture did not give security issues too much attention and impersonation is very likely.

**Method of Preparing the Questions**

The first step in preparing the e-examination questions is to ask the lecturer in-charge of the course to submit the questions to the administrator at the center via the faculty/school exams officer some days before the commencement of the actual exams. The second step is for the administrator (mostly private operator) to enter the pool of questions into the database. The last step is to set the timing for the exams. The implication here is that, when examination questions passes through so many hands it is likely that the questions may leak, especially when a private individual is involved.

**E-Exams Result Presentation/Checking**

In most of the centers visited in this research work, students don't get to see their results immediately after the exams. In some cases, the results may take weeks or even months before it is made available to the students. This violates one of the main essence of introducing e-exams (instant access to results). This may give room for alteration of students result. There is also no room for the users to see the correction of their tests if they so wish.

**METHODOLOGY**

Six Universities that have been engaging electronic examination were participated in this study across the country, where twenty (20) students (15 male, 5 female) were selected from each University for the interview and questionnaire purposes. Also five (5) Lecturers were selected from each University for the interview on the impacts of electronic examination on their students' performance. University of Ilorin, University of Lagos, University of Nigeria Nsuka, Covenant University Ota, Nigeria Open University of Nigeria and Federal University of Technology Minna.

**Interviews and Questionnaire**

The interview was conducted for the students that have undergone e-Examination for post-UTME and internal universities e-Examination. The questionnaires were also distributed to both students and staff members whose students had been evaluated using e-Examination. The questionnaire consisted of 4 essay questions and 20 scaled items concerning the examinees' acceptance of the Secured Electronic Examination (SEE), and its usability. A scale from 1 (total disagreement) to 5 (total agreement) was used. The major aim of the research work is to determine the acceptability or otherwise of the existing electronic system of examining students in our tertiary institution and come out with a new design (Secured e-Examination) based on the deficiency of the existing system

The result of the interview was analyzed in the table below:

| University | Students | | | Lecturer | | |
|---|---|---|---|---|---|---|
| | Good/Acceptance (20) | Fair/Rejection (20) | Indifference(20) | Good/Acceptance (5) | Fair/Rejection (5) | Indifference (5) |
| University of Ilorin | 5 | 15 | | 2 | 3 | |
| FUT Minna | 10 | 10 | | 2 | 2 | 1 |
| Convenant, University Otta | 5 | 15 | | 1 | 4 | |
| University of Nigeria Nnsuka | 5 | 13 | 2 | 1 | 4 | |
| NOUN | 5 | 14 | 1 | 2 | 3 | |
| University of Lagos | 5 | 15 | | 3 | 2 | |
| **Total** | **40** | **77** | **3** | **11** | **18** | **1** |

Table 1: Presentation of the Questionnaire

**Analysis of the Results**

From the table above, it was discovered that out of 120 students that were interviewed, 40 students accept that the present e-Examination is good enough, 77 students reject the use of existing system say that it need to be improved while only 7 students were indifferent to the interview. Also, out of 30 lecturers across the Universities participated in the interview, 11 lecturer attested that existing e-Examination is good enough while 18 lecturers has one or more complains about the system and only one lecturer did not respond to the questions.

The deduction from the table was that the majority of the students and staff prefer e-Examination to manual methods of examining students but want improvement on the system

From the table above:

1. *Acceptance / Good* implies that the existing e-Exam is accepted

2. *Rejection / Fair* implies that the existing e-Exam is totally accepted i.e require further improvement

**CHALLENGES OF THE EXISTING SYSTEM**

- Security
  Both existing biometric and non-biometric e-Examination system involved sending examination questions to the e-Exam centre from the department, where operator will then enter the questions into the system. The biometric system consists of picture box and fingerprint scanner that collect the biometric data of the candidates. But due to the transferring of the question involved, the security of the system is at risk and there is a need to take care of this by designing intranet and send the question through the internet in encrypting language while the questions will be decrypted at the opening of the questions to be answered by the candidates. With the se operators at the centre will not be able to interfere with the questions but just to take care of candidate complains.

- Human interference
  As long as human being is monitoring the e-Exams, it will certainly be influenced by the invigilator. There should be a data capturing and monitoring machine that can revealed the activities of the examinee during the e-Examination. Another area where human can interfere is the delay of the e-Exam result where examinee has to wait for more days to collect his results. The examinee should be able to check their result immediately after the e-Exam so as to prevent the human manipulation of result of whatever kind.

- Inadequate training for the students and Staff
  Many candidates engaging the e-Exam do not understand the proper usage of computer system talk less of the system software. There should be adequate training and awareness for the students prior to the period of e-Examination. So also the staff should be well informed the issues concerning e-Examination.

- Complexity of Software
  The software being used in most of the e-center is a little bit cumbersome. The interface of the software to be used should be highly friendly to increase the effectiveness of the system.

**PROSPECTS**

E-exams simply the process by which examinations are delivered, taken and scored electronically. It entails questions being deployed onto computer workstations (intranet and internet) and candidates answering the questions on to the computer. The process of writing exams is thus completely paperless. It is sometimes referred to as CBT (Computer-based testing) or CBA (Computer-Based Assessment). This testing method is now being extensively used in many parts of the world today. The use of e-exam simplifies the entire testing cycle, including generation, execution, evaluation, presentation and archiving. This simplification saves time and money while improving reliability. Advocates for the e-exams models argue that it not time-consuming but rather time saving, (McCormack and Jones 1998, Ryan et al 2000) and identify these advantages:

- Time saving; as assessments can be created using software tools and adapted and reused as needed. They can be distributed and collected using a web-based system which saves development and distribution time.

- Reduces turnaround time; as the systems enables assessments to be corrected by computers. Reduces time further enables students to use the knowledge obtained from corrected assessments to address further assessments sooner.

- Reduces resources needed by replacing human resources with computer resources.

- Keeping records of results that can be stored centrally and assessed by interested parties, such as students and staff.
- A key element in computer-based testing is that fewer people are required to supervise each examination. This will result in considerable cost savings. While the thought of a computer-based assessment or electronic assessment may intimidate those who are unfamiliar with a computer, electronic tests require only minimal computer knowledge and will offer a familiarity tutorial allowing the test-taker to get acquainted with how to move the mouse, answer questions and move through the test. With computer-based assessment comes the possibility of radically changing how assessments are implemented and improving the quality of the information they can yield.

- Increasing ease with which data can be used as corrected assignments corrected and stored electronically can be analyzed easier and the data can be used in spreadsheets and other statistical packages.
- Now-a- days institutes are organizing exams online. In this module a user can give online exam of a particular subject and get the results instantly through which the user can know his/her potential and how much more effort he/she needs to put in to get better marks. No time is spent on evaluation that means results are available instantly.
- The best available physical and data security techniques to protect the integrity of our tests and to ensure that each candidate takes the exam in a controlled environment. We are proposing stringent security policies and procedures to protect the content of all examinations, ensure that candidate taking the test is the person he/she is supposed to be, ensure that the candidate takes the test unaided and maintain security of data concerning the candidate and the testing session.

**THE PROPOSED SYSTEM**

Secured electronic exams are one of the most difficult challenges in e-learning security. The relevance of the examination process for any academic institution implies that different security mechanisms must be applied in order to preserve some security properties during different examination stages. We are proposing an e-exams system that makes use of both biometric fingerprint authentication and cryptography for protecting questions in order to achieve the desired security levels at every exam stage.

The first stage is that all the principal actors in this system (students, lecturers and administrators) have to register with the biometric fingerprint server as categorized. After the registration, the lecturers can now use their status to send directly the prepared questions and answer keys (encrypted) to the question server (database). The students can now go to the e-exams center and also use their status to login and take the e-exams. The questions are decrypted in the question pool before randomly sent to the student systems. They can easily access their results immediately after the exams. The timing and grading systems are automatic. There is another module that takes care of students complains and correction checking.

Data that can be read and understood without any special measures is called plaintext or cleartext. The method of disguising plaintext in such a way as to hide its substance is called encryption. Encrypting plaintext results in unreadable gibberish called ciphertext. We use encryption to ensure that information is hidden from anyone for whom it is not intended, even those who can see the encrypted data (Huszti and Petho, 2008). The process of reverting ciphertext to its original plaintext is called decryption. We present the proposed system below using the Use-Case diagram.

Use-cases are techniques for capturing the functional requirements of a System. Use-cases work by describing the typical interactions between the users of a System and the System itself, providing a narrative of how a System is used.

47.7

*Figure 1: Use-Case Diagram for E-Exams System*

**The Biometric Fingerprint System**

Biometrics consists of automated methods of recognizing a person based on unique physical characteristic. Each type of biometric system, while different in application, contains at least one similarity: the biometric must be based upon a distinguishable human attribute such as a person's fingerprint, iris, voice pattern or even facial pattern. Fingerprints are the most commonly used biometrics solution as they are less expensive compared with other biometrics solutions.

Biometric fingerprint authentication devices are now available in Nigeria. For example, some laptops come with it and also a biometrics mouse by JayPeetek Inc. called Scan.U.Match is in Nigerian market too. This devise is part of a package of fingerprint authentication mechanism. The mouse is about the same size as standard mouse, however, it also has an integrated fingerprint scanner that is managed by client side software and controlled by server side software centralized on an authentication server.

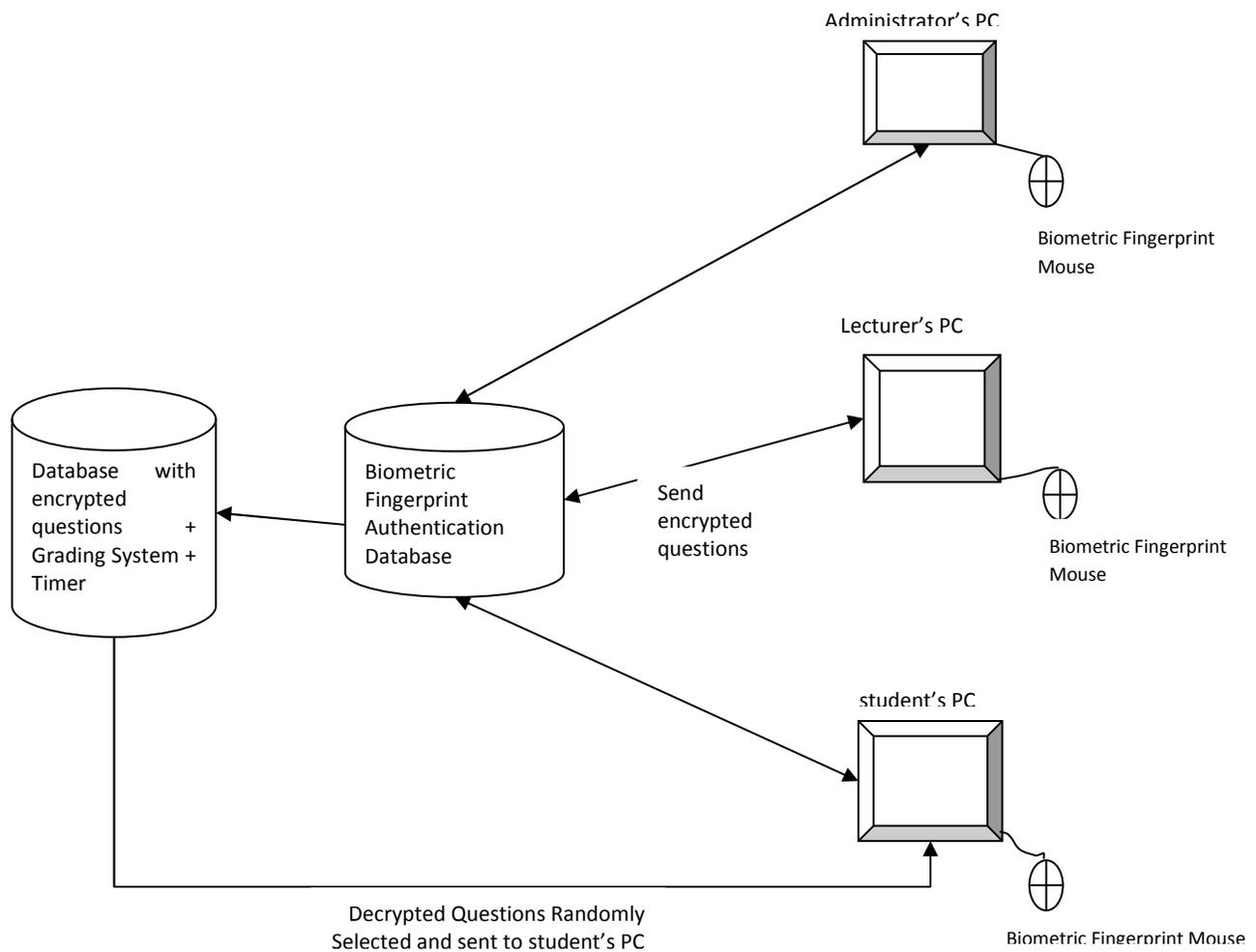

*Figure 2: Proposed Biometric Fingerprint and Cryptographic Solution For E-Exams*

**IMPLEMENTATION AND TESTING**

There are so many programming languages that can be used to implement this proposal. To test the effectiveness of the design, we used Java Applet, PHP and HTML, with MySQL as the back-end integration database. The choice of these programming languages is based on the features of the languages that makes them very appropriate for this work.

We setup a small laboratory containing ten computers and a server to test the workability of the system. The software was tested using four different Web browsers namely Internet Explorer 8, Mozilla Firefox version 3, Netscape Navigator 9.0.0.6 and Opera 9.5. Four different operating systems are also used to test run the software, these are; Microsoft Windows XP, Windows Vista, Windows 7 and Linux (Ubuntu). All the test results are very encouraging and successful.

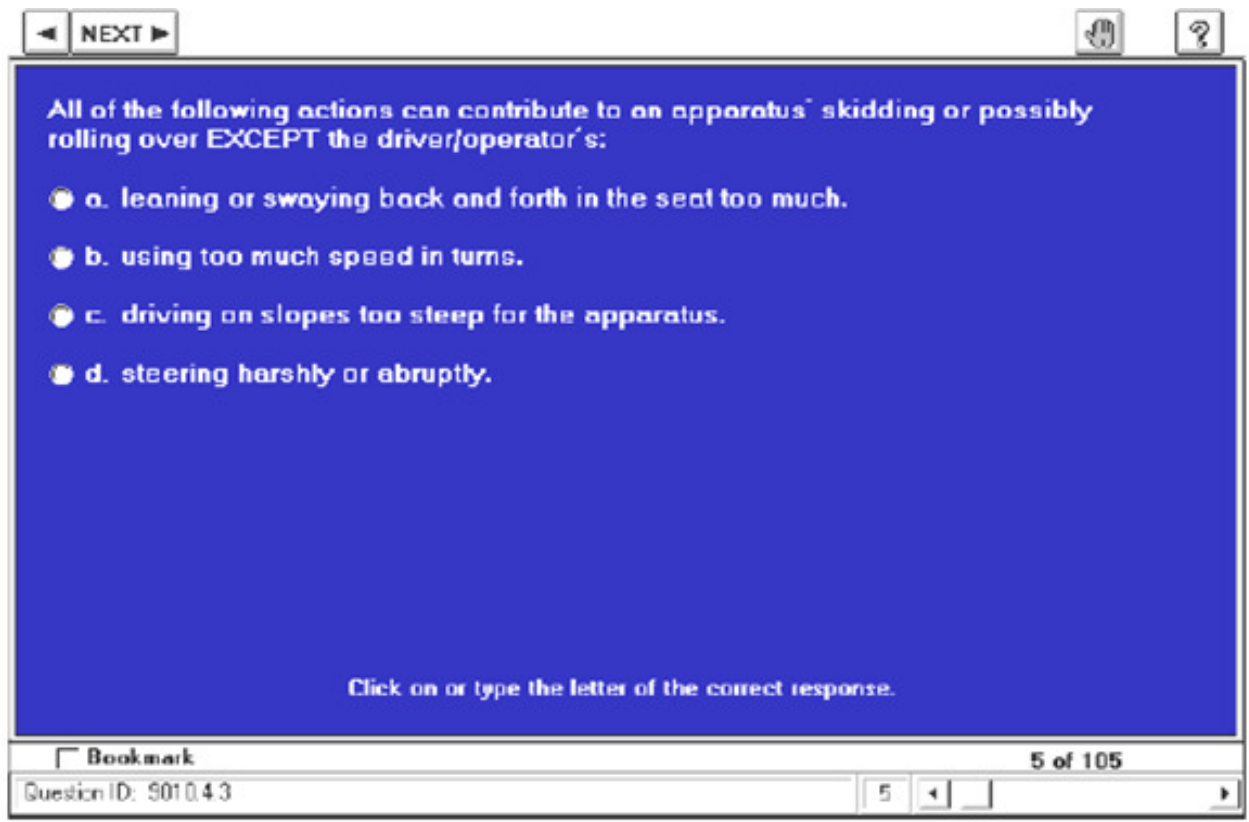

Figure 3: Interface of Multiple Questions E-Exams System

**CONCLUSION**

When fully implemented, the proposed system will definitely reduce drastically the problems mentioned above. These problems includes human interference, impersonation, bribe-taking by lecturers, invigilators and supervisors, too much paper work, examination leakages and also reduce the number of invigilators needed for invigilators. Especially, security will be more effective since the system includes biometric fingerprint authentication, picture capture and data encryption and decryption have been added to the existing design.

Candidates screening is now online and real-time. The system have the tendencies of increasing computer literacy, online learning and network security awareness. The result integrity could also be

47.10

enhanced if the candidates have access to instant result checking. The new system also allows the students to check the corrections at their own wish after the exams.

Future research may be fruitful by examining students' attitudes and psychological aspects associated with the proposed solution of e-exam user's authentication. Furthermore, future research may look at the economical issues associated with implementation of such solution.